\documentclass[twocolumn,floatfix,prl,showkeys]{revtex4}%
\usepackage{graphicx}%
\usepackage{amsmath}%
\usepackage{amsfonts}%
\usepackage{amssymb}
\usepackage{bm}
\usepackage{color}

\def\s{{\sigma}}
\def\e{{\epsilon}}
\def\k{{ {\bm k} }}

\def\q{{ {\bm q} }}
\def\Q{{ {\bm Q} }}
\def\0{{ {\bm 0} }}

\def\w{{\omega}}
\def\a{{\alpha}}

\allowdisplaybreaks[4]

\begin{document}
\title{
Competing Unconventional Charge-Density-Wave States 
in Cuprate Superconductors: \\
Spin-Fluctuation-Driven Mechanism
}
\author{
Kouki Kawaguchi$^{1}$, 
Youichi Yamakawa$^{1}$,
Masahisa Tsuchiizu$^{2}$, and
Hiroshi Kontani$^{1}$
}

\date{\today}

\begin{abstract}
To understand the origin of unconventional
charge-density-wave (CDW) states in cuprate superconductors,
we establish the self-consistent CDW equation, 
and analyze the CDW instabilities based on the realistic Hubbard model,
without assuming any $\q$-dependence and the form factor.
Many higher-order many-body processes, which are
called the vertex corrections, are systematically 
generated by solving the CDW equation.
When the spin fluctuations are strong, the uniform $\q={\bm 0}$
nematic CDW with $d$-form factor shows the leading instability.
The axial nematic CDW instability at $\q=\Q_a=(\delta,0)$ 
($\delta\approx \pi/2$)
is the second strongest, and its strength increases under the 
static uniform CDW order.
The present theory predicts that uniform CDW transition emerges 
at a high temperature,
and it stabilize the axial $\q=\Q_a$ CDW at $T=T_{\rm CDW}$.
It is confirmed that the higher-order Aslamazov-Larkin processes
cause the CDW orders at both $\q={\bm 0}$ and $\Q_a$.

\end{abstract}

\address{
$^1$ Department of Physics, Nagoya University,
Furo-cho, Nagoya 464-8602, Japan. 
\\
$^2$ Department of Physics, Nara Women's University, 
Nara 630-8506, Japan
}
 

\sloppy

\maketitle

\begin{figure}[!htb]
\includegraphics[width=.9\linewidth]{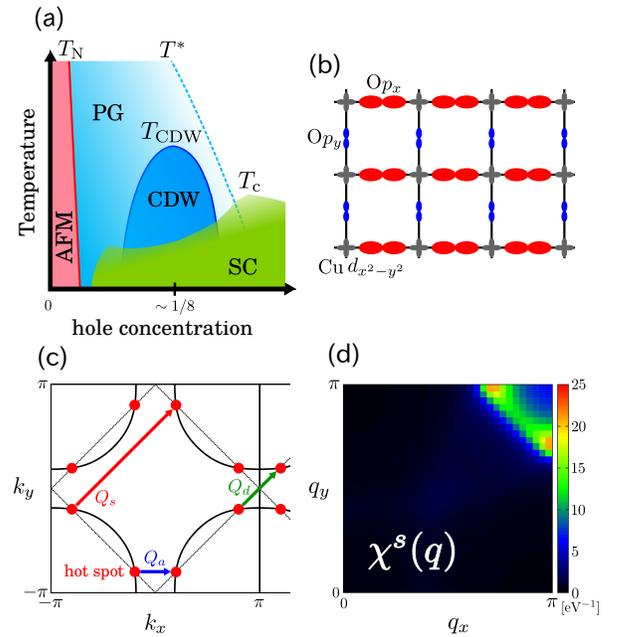}
\caption{(color online)
(a) Schematic phase diagram of hole-doped cuprates
with the pseudogap (PG), charge-density-wave (CDW),
antiferromagnetism (AFM), and superconductivity (SC).
$T^*$ is the pseudogap temperature and
$T_{\rm CDW}$ is the CDW temperature.
(b) $\q={\bm 0}$ CDW due to $p$-orbital polarization ($n_x\ne n_y$)
in real space.
(c) FS of the $d$-$p$ Hubbard model for $n=4.9$.
(d) $\chi^s(\q)$ for $U=4.06$ eV.
}
\label{fig:fig1}
\end{figure}

The origin and the nature of the complex electronic states
in cuprate high-$T_{\rm c}$ superconductors
are central unsolved issues in condensed matter physics.
Recently, interesting interplay between the magnetism, nematicity, 
and superconductivity has been revealed by many experiments.
Figure \ref{fig:fig1} (a) shows the schematic phase diagram
of hole-doped cuprate superconductors.
Strong spin fluctuations develop for wide doping and temperature ranges,
and short-range spin fluctuations are the origin of 
various non-Fermi-liquid-like electronic states
(such as the $T$-linear resistivity and 
$T^{-1}$ behavior of the Hall coefficient)
and the $d$-wave superconductivity at $T_{\rm c}\sim100$ K
 \cite{Moriya,Yamada-text,Scalapino,ROP}.
However, the microscopic origin of the nematicity has been unsolved.
For example, at $T=T_{\rm CDW}$,
the axial charge-density-wave (CDW) at wavevector $\q=\Q_a= (\delta,0)$
($\delta\approx\pi/2$), which is parallel to the nearest Cu-Cu direction,
is observed by the X-ray scattering studies
\cite{Y-Xray1,Y-Xray2,Y-Xray3,Bi-Xray1,Bi-Xray2,Hg-Xray,La-Xray,p-CDW},
STM studies
\cite{STM-Hanaguri,STM-Kohsaka,STM-Lawler,STM-Fujita},
and local lattice deformation 
\cite{Bianconi}.
The band-folding due to the CDW should suppress
the density-of-states and spin fluctuations.

There are many open problems on
the nature of the pseudogap phase below $T^*$, such as 
whether it is a distinct phase or a continuous crossover.
Recently, strong evidences for the nematic transition at $T^*$ 
have been reported by the resonant ultrasound spectroscopy \cite{RUS},
ARPES analysis \cite{ARPES-Science2011}, and 
magnetic torque measurement \cite{Matsuda-torque}.
Fundamental questions for theorists are:
what is the order parameter of the nematic phase below $\sim T^*$,
and why the nematic CDW is realized inside the pseudogap phase.
The mean-field-level approximations, such as the 
random-phase-approximation (RPA),
cannot explain any CDW instabilities unless sizable
inter-site Coulomb interactions are introduced \cite{Bulut,Yamakawa-CDW}.
Thus, we study the role of the vertex corrections (VCs)
that describes the strong charge-spin interference 
\cite{Onari-SCVC,Tsuchiizu-Ru,Onari-SCVCS,Tsuchiizu-CDW,Kontani-Raman}.

\begin{figure}[!htb]
\includegraphics[width=.9\linewidth]{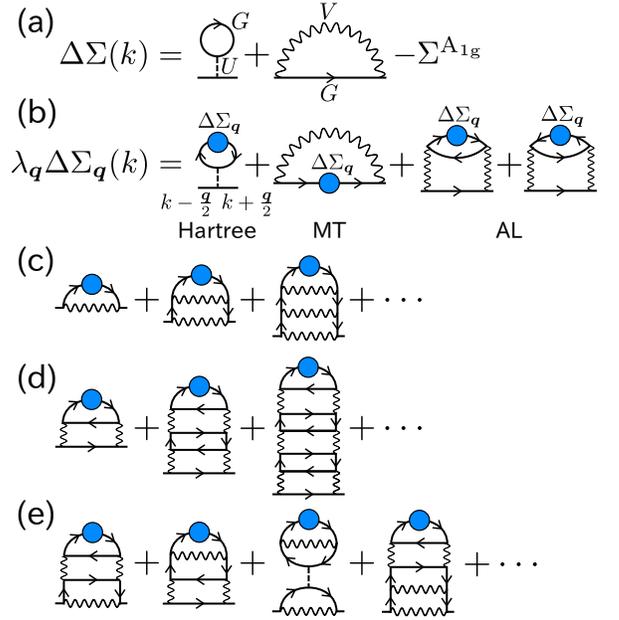}
\caption{(color online)
(a) Schematic self-consistent equation for the CDW order parameter. 
$\Delta\Sigma$ is the symmetry-breaking self-energy 
that represents the CDW order. 
(b) Schematic linearized CDW order equation for general wavenumber $\q$.
(c) Higher-order MT processes.
(d) Higher-order AL processes.
(e) Examples of the mixture of different processes.
}
\label{fig:fig2}
\end{figure}

The idea of the ``spin-fluctuation-driven CDW''
due to the VCs has been studied intensively 
\cite{Onari-SCVC,Tsuchiizu-CDW,Sachdev,Metzner,Chubukov,DHLee-PNAS,Kivelson-NJP}.
By developing this idea,
the electronic nematic phases in Fe-based superconductors \cite{Onari-SCVC},
Ru-oxides \cite{Tsuchiizu-Ru},
and cuprate superconductors 
\cite{Sachdev,Metzner,Chubukov,DHLee-PNAS,Kivelson-NJP,Schmalien-CDW}
have been explained.
The irreducible VC derived from the Ward-identity
($\delta{\hat \Sigma}/\delta {\hat G}$) 
within the one-loop approximation
is given by the single- and double-fluctuation terms, respectively
called the Maki-Thompson (MT) and Aslamazov-Larkin (AL) VCs 
(see Fig. \ref{fig:fig2} (b)).
As studied in Ref. \cite{Sachdev},
the higher-order MT processes give the diagonal CDW with 
$\q=\Q_d=(\delta,\delta)$, which is, however,
inconsistent with experiments.
The axial CDW is given by the lowest-order AL process 
if small inter-site Coulomb interaction exists \cite{Yamakawa-CDW}.
However, the uniform nematic order that corresponds to the pseudogap phase
is failed to be explained.
Therefore, new theoretical method should be developed.

In this paper, 
we establish the self-consistent CDW equation, and 
analyze the CDW instabilities based on the $d$-$p$ Hubbard model.
By solving the CDW equation,
higher-order VCs given by the repetition of the AL and MT processes
are systematically included,
without assuming any $\q$-dependence and the form factor.
When the spin fluctuations are strong, we obtain the uniform CDW 
with $p$-orbital polarization ($n_{x}\ne n_{y}$),
which is schematically shown in Fig. \ref{fig:fig1} (b).
This uniform $p$-orbital polarization strongly enlarges the 
axial nematic CDW instability at $\q=\Q_a$.
The present study leads to the prediction that 
the uniform $p$-orbital polarization occurs at 
$T^*$, and axial $\q=\Q_a$ CDW is induced at $T_{\rm CDW}<T^*$.
We verified that the higher-order AL processes
are significant for the rich CDW orders in under-doped cuprates.

We analyze the three-orbital $d$-$p$ model $H=H_0+H_U$
introduced in Ref. \cite{Yamakawa-CDW}:
The kinetic term is 
$H_0= \sum_{\k,\s} {\hat c}_{\k,\s}^\dagger {\hat h}_\k {\hat c}_{\k,\s}$,
where ${\hat c}_{\k,\s}^\dagger \equiv 
(d_{\k,\s}^\dagger,p_{x,\k,\s}^\dagger,p_{y,\k,\s}^\dagger)$, and
${\hat h}_\k$ is the first-principles tight-binding model 
for La$_2$CuO$_4$ \cite{Held} with the additional 
3rd-nearest $d$-$d$ hopping $t_{dd}^{\rm 3rd}=-0.1$ eV.
The Fermi surface (FS) for the electron filling $n=n_d+n_p=4.9$
shown in Fig. \ref{fig:fig1} (c) is similar to the
hole-like FS in Y- and Bi-based cuprates.
In the interaction term $H_U$,
we introduce only the $d$-orbital on-site Coulomb interactions $U$.
In the RPA, the spin (charge) susceptibility for the $d$-orbital is 
${\chi}^{s(c)}(q)={\chi}^{0}(q)/({1}-(+){U}{\chi}^{0}(q))$, 
where $q\equiv(\q,\w_l)$ and $\w_l=2l\pi T$.
Here, $\chi^{0}(q)=-T\sum_{k}G(k+q)G(k)$ is the bare bubble, 
and $G(k)= ({\hat 1}(i\e_n-\mu)-\Delta{\hat \Sigma}(k)-{\hat h}_\k)^{-1}_{1,1}$ 
is the Green function for the $d$-orbital.
Here, $k\equiv(\k,\e_n)$, $\e_n=(2n+1)\pi T$,
and $\Delta\Sigma(k)$ is the symmetry-breaking self-energy;
we will introduce later.
Figure \ref{fig:fig1} (d) shows the obtained spin susceptibility
for $\Delta\Sigma(k)=0$, in the case of $U=4.06$ eV, $n=4.9$ and $T=50$ meV.
The spin Stoner factor $\a_S\equiv \max_\q\{U{\chi}^{0}(\q)\}$ is 0.99.
Hereafter, we fix the parameters $n=4.9$ and $T=50$ meV.

In principle, the CDW order parameter is given as the 
``symmetry breaking in the self-energy'',
similarly to the superconductivity given as the
symmetry-breaking in the anomalous self-energy.
In Ref. \cite{Onari-FeSe}, we developed the symmetry-breaking 
self-energy method, and explained the nematic orbital order
in Fe-based superconductors.
The non-magnetic nematic order with the  
sign-reversing orbital polarization in $\k$-space in FeSe 
has been satisfactorily explained \cite{Onari-FeSe}.
Here, we apply the same method to analyze the CDW in cuprates.
The symmetry-breaking self-energy equation is given as
\begin{eqnarray}
\Delta\Sigma(k)&=& (1-P_{\rm A_{1g}})T\sum_q 
\left(\frac32 V^s(q)+\frac12 V^c(q)-U^2\chi^0(q) \right)
\nonumber \\
& &\times G(k+q),
\label{eqn:self}
\end{eqnarray}
where 
$V^s(q)=U+U^2\chi^s(q)$, $V^c(q)=-U+U^2\chi^c(q)$, 
and $G(k)$ contains the symmetry-breaking term $\Delta \Sigma$.
$P_{\rm A_{1g}}$ is the ${\rm A}_{\rm 1g}$ symmetry projection operator.
This equation is shown in Fig. \ref{fig:fig2} (a).
In the FeSe model, the nonzero solution $\Delta \Sigma\ne0$ 
with the ${\rm B}_{\rm 1g}$ symmetry at $\q={\bm 0}$ 
is obtained for $\a_S\ge0.82$ 
\cite{Onari-FeSe}.
In order to study the CDW at $\q\ne{\bm 0}$, however,
we have to calculate Eq. (\ref{eqn:self})
for the large cluster model.

In order to analyze the CDW state with arbitrary wavevector $\q$,
we introduce the linearized CDW equation 
by linearizing Eq. (\ref{eqn:self}) with respect to $\Delta\Sigma$.
The obtained equation is
\begin{eqnarray}
&& \lambda_\q \Delta\Sigma_\q(k)= T\sum_{k'} K(\q;k,k')\Delta\Sigma_\q(k'),
\label{eqn:linearized}    
\end{eqnarray}
where $\lambda_\q$ is the eigenvalue for the CDW at $\q$.
When the maximum of $\lambda_\q$ reaches unity at $\q$,
Eq. (\ref{eqn:self}) for the sufficiently large cluster model 
has nonzero solution, and 
the eigenvector $\Delta\Sigma_\q(k)$ gives the CDW form factor.
The kernel $K(\q;k,k')$ is given as
\begin{eqnarray}
&& \!\!\!\!\!\!\!\!\!\!\!
K(\q;k,k')=
\left(\frac32 V_0^s(k-k')+\frac12 V_0^c(k-k')\right)
\nonumber \\
&& \ \ \ \ \ \ \ \ \ \ 
\times G_0(k'+{\q}/{2})G_0(k'-{\q}/{2})
 \nonumber \\
&&  \ \ \ \
-T\sum_p \left( \frac32 V_0^s(p+{\q}/{2})V_0^s(p-{\q}/{2}) \right.
 \nonumber \\
&& \ \ \ \ \ \ \ \
\left. +\frac12 V_0^c(p+{\q}/{2})V_0^c(p-{\q}/{2}) \right)
\nonumber \\
&& \ \ \ \ \ \ \ \ \ \ 
\times G_0(k-p) \left(\Lambda_\q(k';p)+\Lambda_\q(k';-p)\right) ,
\label{eqn:K} 
\end{eqnarray}
where $\Lambda_\q(k;p)\equiv G_0(k+\frac{\q}{2})G_0(k-\frac{\q}{2})G_0(k-p)$.
The subscript 0 in Eq. (\ref{eqn:K}) represents the 
functions with $\Delta\Sigma=0$.
(In Eq. (\ref{eqn:K}), we should subtract the double-counting term 
in the $U^2$-order.)
The diagrammatic expression for Eqs. (\ref{eqn:linearized})-(\ref{eqn:K})
is shown in Fig. \ref{fig:fig2} (b):
The kernel $K(\q;k,k')$ contains the Hartree term, 
the MT term, and the two AL terms.
Both the MT term and the AL terms may drive the spin-fluctuation-driven CDW.
The AL terms drives the $\q={\bm0}$ CDW instability 
since its functional form $\propto \sum_\k\chi^s(\k+\q)\chi^s(\k)$ 
is large for $\q\approx{\bm0}$
\cite{Onari-SCVC,Grilli}.
Although the Hartree term suppress the CDW instability 
in single-orbital models, its suppression disappears
if the form factor has sign-reversal.

By solving the linearized equation,
the higher-order diagrams with respect to these terms are generated.
For instance, we show the higher-order MT and AL processes 
in Figs. \ref{fig:fig2} (c) and (d), respectively.
The examples of the mixture of different processes are 
given in Fig. \ref{fig:fig2} (e).

Hereafter, we analyze the linearized CDW equation 
in Eq. (\ref{eqn:linearized}) numerically.
Here, we drop the $\e_n$-dependence of $\Delta\Sigma_\q(k)$
by performing the analytic continuation 
$i\e_n \rightarrow \e$ and putting $\e=0$.
We also neglect the $\e_n$-dependent self-energy 
due to spin fluctuations in the kernel $K(\q;k,k')$.
Due to these simplifications, 
the obtained $\lambda_\q$ is expected to be overestimated.
Therefore, we do not put the constraint $\lambda_\q<1$
in the numerical study.
We will show below that $\lambda_\q$ is actually reduced by 
introducing the constant damping $\gamma$
in Green functions (not in $V_0^{s,c}(q)$) in Eq. (\ref{eqn:K}).

\begin{figure}[!htb]
\includegraphics[width=.9\linewidth]{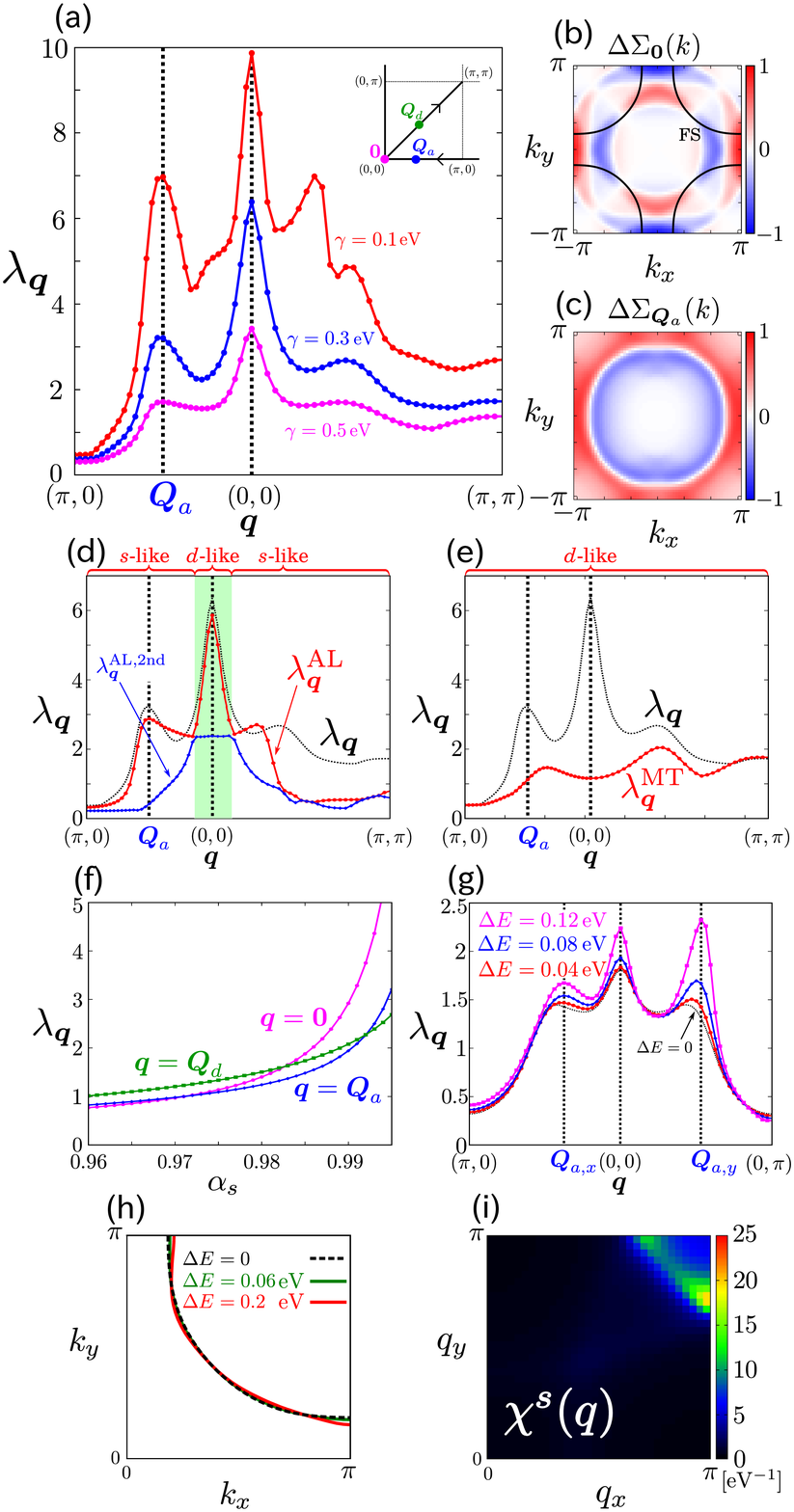}
\caption{(color online)
(a) $\q$-dependence of $\lambda_\q$ obtained for $\gamma=0.1\sim0.5$ eV.
(b) Form factor for $\q={\bm 0}$ ($d$-wave)
noramlized by its maximum value.
(c) Form factor for $\q=\Q_a=(\delta,0)$ ($s$-wave like).
(d) $\lambda_\q^{\rm AL}$ given by including only the AL processes for $\gamma=0.3$ eV .
$\lambda_\q^{\rm AL,2nd}$ is the second largest eigenvalue. 
(e) $\lambda_\q^{\rm MT}$ given by including only the MT processes.
(f) $\lambda_\q$ at $\q=$ $\bm0$, $\Q_a$ and $\Q_d$ as function of $\a_S$.
(g) $\lambda_\q$ under the uniform nematic CDW order
for $\Delta E=0\sim0.12$eV. 
(h) The deformed FS for $\Delta E=0.06$ eV and 0.2 eV.
(i) $\chi^s(\q)$ for $\Delta E=0.06$eV. 
}
\label{fig:fig3}
\end{figure}

Figure \ref{fig:fig3} (a) shows the 
$\q$-dependence of the eigenvalue $\lambda_\q$ obtained for 
$\gamma=0.1\sim0.5$ eV when $\a_S=0.995$ at $T=50$meV.
Here, $\lambda_\q$ is the largest at $\q={\bm 0}$,
meaning that the uniform CDW emerges at the highest temperature.
As shown in Fig. \ref{fig:fig3} (b), the corresponding form factor
$\Delta\Sigma_{{\bm 0}}(\k)$ has the $d$-wave symmetry,
resulting in the uniform $p$-orbital polarization with $n_{px}\ne n_{py}$
shown in Fig. \ref{fig:fig1} (b).
The second largest peak appears at $\q=\Q_a=(\delta,0)$,
which corresponds to the axial CDW.
Its form factor $\Delta\Sigma_{\Q_a}(\k)$ is $s$-wave like,
shown in Fig. \ref{fig:fig3} (c).
Since these form factors have sign reversal in $\k$-space,
the contribution from the Hartree term in Eq. (\ref{eqn:K})
(Fig. \ref{fig:fig2} (b)) is absent or negligibly small.
That is, the present CDW fluctuations driven by the VCs
originate from the irreducible part of the charge susceptibility
with respect to $U$, since the reducible part is negligible
due to the sign-reversing form factor.

To find the origin of the CDW instability,
we solve the linearized CDW equation by including only two AL terms
(MT term) in $K(\q;k,k')$; see Fig. \ref{fig:fig2} (b),
and denote the obtained eigenvalue as $\lambda_\q^{\rm AL(MT)}$.
In Fig. \ref{fig:fig3} (d), we show $\lambda^{\rm AL}_\q$
(and the second-largest eigenvalue $\lambda^{\rm AL, 2nd}_\q$),
for $\a_S=0.995$ and $\gamma=0.3$eV.
At $\q={\bm0}$ and $\Q_a$,
$\lambda^{\rm AL}_\q$ is almost equal to 
the true eigenvalue $\lambda_\q$ shown by the broken line.
The form factor for $\lambda_\q^{\rm AL}$ is 
$d$-wave only for $\q\approx{\bm0}$, shown by the shaded area.
Outside this area, the $d$-wave eigenvalue decreases 
and replaced with the $s$-wave solution.
In contrast, $\lambda^{\rm MT}_\q$ shown in Fig. \ref{fig:fig3} (e)
is much smaller than $\lambda_\q$ at $\q={\bm0}$ and $\Q_a$,
whereas  $\lambda_\q$ at $\q=\Q_d=(\delta,\delta)$ 
is comparable to the true eigenvalue.
Therefore, the CDW instabilities at $\q={\bm0}$ and $\Q_a$
originate from the AL processes, whereas the instability at $\q=\Q_d$ 
is mainly derived from the MT processes.

Figure \ref{fig:fig3} (f) shows the eigenvalues 
at $\q={\bm 0}$, $\Q_a$, and $\Q_d$ as function of $\a_S$.
When the spin fluctuations are smaller 
($\a_S\lesssim0.98$),
$\lambda_\q$ has the maximum at $\q=\Q_d$,
since the spin-fluctuation-driven VCs are small.
As the spin fluctuations increases,
$\lambda_\q$ is drastically enlarged by the VCs,
and $\lambda_{{\bm0}}$ becomes the largest due to the AL processes.
The relation $\lambda_{\Q_a}>\lambda_{\Q_d}$
is realized for $\a_S\gtrsim0.99$.

Hereafter, we study the linearized CDW equation 
under the static uniform CDW order;
$\Delta E(\k)=\Delta E\times \Delta{\Sigma}_{\bm0}(\k)$,
where $\Delta{\Sigma}_{\bm0}(\k)$ 
is normalized by its maximum value.
That is, ${\rm max}_\k \{\Delta E(\k)\}=\Delta E$.
Figure \ref{fig:fig3} (g) shows the eigenvalues 
for $\Delta E=0\sim0.12$ eV at $U=4.04$eV.
For $\Delta E\ge0.1$eV, $\lambda_{\q}$ at $\q=\Q_{a,y}=(0,\delta)$
is larger than the eigenvalues at $\q={\bm0}$ and 
$\Q_{a,x}=(\delta,0)$.
Thus, the axial CDW instability is magnified by the 
nematic FS deformation under the uniform CDW order.
Figures \ref{fig:fig3} (h) and (i) show the $C_2$-symmetric FSs 
and $\chi^s(\q)$, respectively.
For $\Delta E=0.06$ eV, 
the change in the filling is $(\Delta n_d, \Delta n_{p_x}, \Delta n_{p_y})
=(-0.034,-0.179,+0.213)\times 10^{-3}$.
The prominent $C_2$-symmetric spin susceptibility
is consistent with the neutron study for under-doped YBCO 
\cite{YBCO-neutron-C2}.

\begin{figure}[!htb]
\includegraphics[width=.7\linewidth]{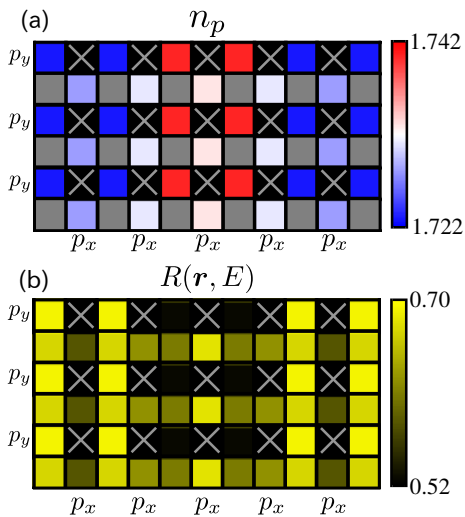}
\caption{(color online)
(a) $\Delta n_{px}$ and $\Delta n_{py}$ on O sites in the CDW state
given by the form factor at $\q=(\pi/2,0)$ for $\Delta E=0.4$eV.
No atoms exist at the cross sites.
(b) $R({\bm r},E)=I({\bm r},E)/I({\bm r},-E)$ for $E=0.5$ eV
at Cu and O sites,
which is similar to the STM data \cite{STM-Kohsaka,STM-Fujita}.
}
\label{fig:fig4}
\end{figure}

The form factor $\Delta \Sigma_\q(\k)$
gives the modulation of the charge density in real space.
Since the $d$-wave form factor at $\q={\bm 0}$ represents 
the bond order $t_{dd}^x \ne t_{dd}^y$, it gives the
orbital polarization $n_{px} \ne n_{py}$ shown in Fig. \ref{fig:fig1} (b).
Figure \ref{fig:fig4} (a) shows the $p$-orbital polarization 
derived from the axial CDW form factor 
$\Delta E(\k)\propto \Delta \Sigma_{\Q_{a,x}}(\k)$ with $\Delta E=0.4$ eV.
Although the form factor in Fig. \ref{fig:fig3} (c) is $s$-wave like,
the obtained $p$-orbital polarization is antiphase 
between the nearest $p_x$ and $p_y$ sites ($d$-wave type).
We also calculate the ratio
$R({\bm r},E)=I({\bm r},E)/I({\bm r},-E)$,
where $I({\bm r},E)=\int_0^E N({\bm r},E')dE'$ 
is the tunneling current in the STM study.
The obtained result for $E=0.5$eV is shown in Fig. \ref{fig:fig4} (b),
which reproduces the characteristic STM pattern reported in 
Refs. \cite{STM-Kohsaka,STM-Fujita}.

In Ref. \cite{Yamakawa-CDW},
we have shown that the axial CDW order at $\q=(\delta,0)$,
by considering the lowest-order AL process and introducing 
small inter-site Coulomb interaction $V_{dp}$.
The obtained $\delta$ decreases with the hole-doping 
in proportion to the wavelength between the hot spots \cite{Yamakawa-CDW}.
On the other hand, the axial CDW order is obtained even if $V_{dp}=0$ 
by the functional-renormalization-group (fRG) analysis
\cite{Tsuchiizu-CDW}.
The preset study clarified that the 
higher-order AL processes in Fig. \ref{fig:fig2} (d)
causes divergent CDW instabilities at $\q={\bm0}$ and $\Q_a$ 
even for $V_{dp}=0$.
Recently, we improved the numerical accuracy of the fRG method 
at $\q\sim{\bm0}$, and verified that the uniform CDW instability 
is the strongest \cite{Tsuchiizu-future}, 
consistently with the present theory.

The pseudogap phenomena under $T^*$,
such as the Fermi arc formation 
\cite{Arc-Yoshida,Arc-Kanigel,Arc-Kondo},
remains unsolved since the uniform CDW cannot account for the pseudogap.
We consider that the short-range spin-fluctuations at $T\sim T^*$
induces not only the uniform CDW due to the AL processes,
but also the large quasiparticle damping \cite{TPSC,Kotliar,Maier},
and the latter causes the pseudogap behaviors.
It is also our important future issue to
explain the doping dependence of $T^*$, $T_{\rm CDW}$ and $T_{\rm c}$ 
quantitatively by improving the theoretical method.

In summary,
we analyzed the linearized CDW equation
based on the $d$-$p$ Hubbard model, 
by including both the MT and AL VCs into the kernel.
When the spin fluctuations are strong ($\a_S\gtrsim0.98$),
the uniform nematic CDW has the strongest instability.
The axial CDW instability is strongly magnified 
under the uniform CDW order, even if the deformation of the FS is small.
These results lead to the prediction that 
the uniform $p$-orbital polarization occurs at
$\sim T^*$, and axial $\q=\Q_a$ CDW is shown in Fig. \ref{fig:fig1} (a) is
induced at $T_{\rm CDW}<T^*$.\
We verified that the higher-order AL processes
give the rich CDW orders.
Various rich spin-fluctuation-driven charge orders
(such as the CDW and orbital order)
are caused by the VCs not only in cuprate and Fe-based superconductors,
but also other metals near the magnetic criticality.

\acknowledgements
We are grateful to S. Onari, Y. Matsuda, T. Hanaguri, 
T Shibauchi, Y. Kasahara, and Y. Gallais
for fruitful discussions.
This study has been supported by Grants-in-Aid for Scientific 
Research from MEXT of Japan.



\begin{thebibliography}{99}

\bibitem{Moriya}
T. Moriya and K. Ueda, Adv. Phys. {\bf 49}, 555 (2000).

\bibitem{Yamada-text}
K. Yamada: {\it Electron Correlation in Metals} 
(Cambridge Univ. Press 2004).

\bibitem{Scalapino}
D. J. Scalapino, Phys. Rep. {\bf 250}, 329 (1995).

\bibitem{ROP}
H. Kontani, Rep. Prog. Phys. {\bf 71}, 026501 (2008);
H. Kontani, {\it Transport Phenomena in Strongly Correlated Fermi Liquids}
(Springer-Verlag Berlin and Heidelberg GmbH \& Co. K, 2013).

\bibitem{Y-Xray1}
G. Ghiringhelli, M. L. Tacon, M. Minola, S. Blanco-Canosa, C. Mazzoli, 
N. B. Brookes, G. M. D. Luca, A. Frano, D. G. Hawthorn, F. He, T. Loew, 
M. M. Sala, D. C. Peets, M. Salluzzo, E. Schierle, R. Sutarto, 
G. A. Sawatzky, E. Weschke, B. Keimer, and L. Braicovich, 
Science {\bf 337}, 821 (2012).

\bibitem{Y-Xray2}
J. Chang, E. Blackburn, A. T. Holmes, N. B. Christensen, J. Larsen, J. Mesot, 
R. Liang, D. A. Bonn, W. N. Hardy, A. Watenphul, M. v. Zimmermann, 
E. M. Forgan, and S. M. Hayden, 
Nature Physics {\bf 8}, 871 (2012).

\bibitem{Y-Xray3}
E. Blackburn, J. Chang, M. H\"{u}cker, A. T. Holmes, N. B. Christensen, 
R. Liang, D. A. Bonn, W. N. Hardy, U. R\"{u}tt, O. Gutowski, 
M. v. Zimmermann, E. M. Forgan, and S. M. Hayden, 
Phys. Rev. Lett. {\bf 110}, 137004 (2013).


\bibitem{Bi-Xray1}
R. Comin, A. Frano, M. M. Yee, Y. Yoshida, H. Eisaki, E. Schierle,
E. Weschke, R. Sutarto, F. He, A. Soumyanarayanan, Y. He, 
M. L. Tacon, I. S. Elfimov, J. E. Hoffman, G. A. Sawatzky, 
B. Keimer, and A. Damascelli,
Science {\bf 343}, 390 (2014).

\bibitem{Bi-Xray2}
E. H. da Silva Neto, P. Aynajian, A. Frano, R. Comin,
E. Schierle, E. Weschke, A. Gyenis, J. Wen, J. Schneeloch,
Z. Xu, S. Ono, G. Gu, M. L. Tacon, and A. Yazdani,
Science {\bf 343}, 393 (2014).

 \bibitem{Hg-Xray}
W. Tabis, Y. Li, M. L. Tacon, L. Braicovich, A. Kreyssig, 
M. Minola, G. Dellea, E. Weschke, M. J. Veit, M. Ramazanoglu, 
A. I. Goldman, T. Schmitt, G. Ghiringhelli, N. Bari\u{s}i\'{c}, 
M. K. Chan, C. J. Dorow, G. Yu, X. Zhao, B. Keimer, and M. Greven, 
Nature Commun. {\bf 5}, 5875 (2014).

\bibitem{La-Xray}
M. H\"{u}cker, M. v. Zimmermann, G. D. Gu, Z. J. Xu, J. S. Wen, 
G. Xu, H. J. Kang, A. Zheludev, and J. M. Tranquada, 
Phys. Rev. B {\bf 83}, 104506 (2011).

\bibitem{p-CDW}
R. Comin, R. Sutarto, F. He, E. da Silva Neto, L. Chauviere, A. Frano, 
R. Liang, W. N. Hardy, D. Bonn, Y. Yoshida, H. Eisaki, J. E. Hoffman, 
B. Keimer, G. A. Sawatzky, and A. Damascelli,
Nature Materials {\bf 14}, 796 (2015).

\bibitem{STM-Hanaguri}
T. Hanaguri, C. Lupien, Y. Kohsaka, D.-H. Lee, M. Azuma, M. Takano, 
H. Takagi, and J. C. Davis, Nature {\bf 430}, 1001 (2004).

\bibitem{STM-Kohsaka}
Y. Kohsaka, T. Hanaguri, M. Azuma, M. Takano, J. C. Davis, and H. Takagi,
Nature Physics {\bf 8}, 534 (2012). 

\bibitem{STM-Lawler}
M. J. Lawler, K. Fujita, J. Lee, A. R. Schmidt, Y. Kohsaka, 
C. K. Kim, H. Eisaki, S. Uchida, J. C. Davis, J. P. Sethna, and E.-A. Kim, 
Nature {\bf 466}, 347 (2010).

\bibitem{STM-Fujita}
K. Fujita, M. H. Hamidian, S. D. Edkins, C. K. Kim, Y. Kohsaka, 
M. Azuma, M. Takano, H. Takagi, H. Eisaki, S. Uchida, A. Allais, 
M. J. Lawler, E.-A. Kim, S. Sachdev, and J. C. S. Davis, 
Proc. Natl. Acad. Sci. USA, {\bf 110}, E3026 (2014).

\bibitem{Bianconi}
A. Bianconi, N. L. Saini, A. Lanzara, M. Missori, and T. Rossetti,
H. Oyanagi, H. Yamaguchi, K. Oka, and T. Ito,
Phys. Rev. Lett. {\bf 76}, 3412 (1996).

\bibitem{RUS}
A.Shekhter, B. J. Ramshaw, R. Liang, W. N. Hardy, D. A. Bonn,
F. F. Balakirev, R. D. McDonald, J. B. Betts, S. C. Riggs, and A. Migliori,
Nature {\bf 498}, 75 (2013).

\bibitem{ARPES-Science2011}
Rui-Hua He, M. Hashimoto, H. Karapetyan, J. D. Koralek, J. P. Hinton, J. P. Testaud, V. Nathan, Y. Yoshida, Hong Yao, K. Tanaka, W. Meevasana, R. G. Moore, D. H. Lu, S.-K. Mo, M. Ishikado, H. Eisaki, Z. Hussain, T. P. Devereaux, S. A. Kivelson1, J. Orenstein, A. Kapitulnik, and Z.-X. Shen,
Science {\bf 331}, 1579 (2011).

\bibitem{Matsuda-torque}
Y. Sato, Y. Kasahara, T. Shibauchi and Y. Matsuda, private communication

\bibitem{Bulut}
S. Bulut, W.A. Atkinson and A.P. Kampf,
Phys. Rev. B {\bf 88}, 155132 (2013).

\bibitem{Yamakawa-CDW}
Y. Yamakawa, and H. Kontani,
Phys. Rev. Lett. {\bf 114}, 257001 (2015).

\bibitem{Onari-SCVC}
S. Onari and H. Kontani, 
Phys. Rev. Lett. {\bf 109}, 137001 (2012).


\bibitem{Tsuchiizu-Ru}
M. Tsuchiizu, Y. Ohno, S. Onari, and H. Kontani, 
Phys. Rev. Lett. {\bf 111}, 057003 (2013).


\bibitem{Onari-SCVCS}
S. Onari, Y. Yamakawa, and H. Kontani, 
Phys. Rev. Lett. {\bf 112}, 187001 (2014).

\bibitem{Kontani-Raman}
H. Kontani and Y. Yamakawa, 
Phys. Rev. Lett. {\bf 113}, 047001 (2014).

\bibitem{Tsuchiizu-CDW}
M. Tsuchiizu, Y. Yamakawa, and H. Kontani,
Phys. Rev. B {\bf 93}, 155148 (2016).

\bibitem{Sachdev}
M.A. Metlitski and S. Sachdev, New J. Phys. {\bf 12}, 105007 (2010);
S. Sachdev and R. La Placa, Phys. Rev. Lett. {\bf 111}, 027202 (2013).

\bibitem{Metzner}
C. Husemann and W. Metzner, Phys. Rev. B {\bf 86}, 085113 (2012);
T. Holder and W. Metzner, Phys. Rev. B {\bf 85}, 165130 (2012).

\bibitem{Chubukov}
Y. Wang and A.V. Chubukov, 
Phys. Rev. B {\bf 90}, 035149 (2014).

\bibitem{DHLee-PNAS}
J. C. S. Davis and D.-H. Lee, 
Proc. Natl. Acad. Sci. USA, {\bf 110}, 17623 (2013).

\bibitem{Kivelson-NJP}
E. Berg, E. Fradkin, S. A. Kivelson, and J. M. Tranquada, 
New J. Phys. {\bf 11}, 115004 (2009).

\bibitem{Schmalien-CDW}
P. P. Orth, B. Jeevanesan, R. M. Fernandes, and J. Schmalian,
arXiv:1703.02210.

\bibitem{Held}
P. Hansmann, N. Parragh, A. Toschi, G. Sangiovanni, and K. Held, 
New J. Phys. {\bf 16}, 033009 (2014).

\bibitem{Onari-FeSe}
S. Onari, Y. Yamakawa, and H. Kontani,
Phys. Rev. Lett. {\bf 116}, 227001 (2016).

\bibitem{Grilli}
S. Caprara, C. Di Castro, M. Grilli, and D. Suppa, 
Phys. Rev. Lett. {\bf 95}, 117004 (2005).

\bibitem{YBCO-neutron-C2}
V. Hinkov, D. Haug, B. Fauque, P. Bourges, Y. Sidis, A. Ivanov, C. Bernhard, C. T. Lin, and B. Keimer,
Science, {\bf 319}, 597 (2008).

\bibitem{Tsuchiizu-future}
M. Tsuchiizu, K. Kawaguchi, Y. Yamakawa, and H. Kontani, unpublished.

\bibitem{Arc-Yoshida}
T. Yoshida, X. J. Zhou, T. Sasagawa, W. L. Yang, P. V. Bogdanov, 
A. Lanzara, Z. Hussain, T. Mizokawa, A. Fujimori, H. Eisaki, 
Z.-X. Shen, T. Kakeshita, and S. Uchida,
Phys. Rev. Lett. {\bf 91}, 027001 (2003);
M. Hashimoto, I. M. Vishik, Z.-X. Shen, and A. Fujimori, 
J. Phys. Soc. Jpn. {\bf 81}, 011006 (2012).

\bibitem{Arc-Kanigel}
A. Kanigel, M. R. Norman, M. Randeria, U. Chatterjee, S. Souma, 
A. Kaminski, H. M. Fretwell, S. Rosenkranz, M. Shi, T. Sato, 
T. Takahashi, Z. Z. Li, H. Raffy, K. Kadowaki, D. Hinks,
L. Ozyuzer and J. C. Campuzano,
Nature Physics {\bf 2}, 447 (2006).

\bibitem{Arc-Kondo}
T. Kondo, Y. Hamaya, A. D. Palczewski, T. Takeuchi, J. S. Wen, Z. J. Xu, 
G. Gu, J. Schmalian, and A. Kaminski,
Nature Physics {\bf 7}, 21 (2011).


\bibitem{TPSC}
D. Senechal and A.-M.S. Tremblay, Phys. Rev. Lett. {\bf 92}, 126401 (2004).

\bibitem{Kotliar}
B. Kyung, S. S. Kancharla, D. Senechal, A. -M. S. Tremblay, M. Civelli, and G. Kotliar,
Phys. Rev. B {\bf 73}, 165114 (2006).

\bibitem{Maier}
T.A. Maier, M.S. Jarrell, and D.J. Scalapino,
Physica C, {\bf 460-462}, 13 (2007).


\end{thebibliography}
\end{document}